\documentclass[11pt, preprint]{article}
\usepackage{color} \usepackage{cancel}
\usepackage{float} \usepackage{graphicx}
\usepackage{subfigure} \usepackage{caption}	
\usepackage{dcolumn}
\usepackage[toc,page]{appendix} \usepackage{authblk}	 		
\usepackage{indentfirst}	
\usepackage{authblk}	 		
\usepackage{multirow}     
\usepackage{hyperref}     
\usepackage{amsfonts, amssymb, amsmath}
\usepackage[left=2.5cm,right=2.5cm,top=2.9cm,bottom=2.9cm,a4paper]{geometry}
\usepackage[noadjust]{cite}
\usepackage{filecontents}
\def\ear{\end{eqnarray}}
\def\beq{\begin{equation}}             \def\earn{\nonumber \end{eqnarray}}
\def\eeq{\end{equation}}               
\def\bear{\begin{eqnarray}}

\usepackage{soul}
\usepackage{xcolor}
\setstcolor{red}

\graphicspath{{./figures/}}
\hypersetup{ colorlinks   = true, 
urlcolor     = blue, 
linkcolor    = blue, 
citecolor    = red 	 
}

\title{\Large\bf Spherical symmetric solutions of $f(R)$ gravity \\ with a kinetic curvature scalar}
\author[1,2,3]{S. V. Chervon\thanks{chervon.sergey@gmail.com}}
\author[4,5]{J. C. Fabris\thanks{julio.fabris@cosmo-ufes.org}}
\author[2]{I. V. Fomin\thanks{ingvor@inbox.ru}}
\affil[1]{\small \it Ulyanovsk State Pedagogical University, Lenin's square 4/5, Ulyanovsk 432071, Russia}
\affil[2]{\small \it  Bauman Moscow State Technical University, 2-nd Baumanskaya street, 5, Moscow, 105005, Russia}
\affil[3]{\small \it Kazan Federal University, Kremlevskaya street 18, Kazan, 420008, Russia}
\affil[4]{\small \it COSMO-UFES \& Departamento de Fisica, Universidade Federal do Espirito Santo (UFES)
Av. Fernando Ferrari s/n CEP 29.075-910, Vitoria, ES, Brazil}
\affil[5]{\small \it National Research Nuclear University, MEPhI, Kashirskoe shosse 31,
115409, Moscow, Russia}

\begin{document}

\maketitle
\begin{abstract}
We consider modified $f(R)$ gravity with a kinetic curvature scalar as a
chiral self-gravitating model in a spherically symmetric spacetime. Most attention devoted to finding solutions for special case of scaling transformation when modified gravity transforms to Einstein frame from Jordan one. We proposed the method of determination of kinetic function for given scalar field dependence on space coordinate. New classes of solutions are found for special choice of $f(R) $ function.
\end{abstract}

\bigskip


\section{Introduction}

 The fact that Universe expansion is accelerating at the present time is confirmed by various observations such as measurements from supernovae \cite{cfo-Perlmutter:1998np}, \cite{cfo-Riess:1998cb}, cosmic microwave background (CMB) radiation \cite{Ade:2015xua}, \cite{BICEP2-2015}, \cite{WMAP2013}, large scale structure \cite{Seljak:2004xh}, baryon acoustic oscillation \cite{Eisenstein:2005su} and weak lensing \cite{Jain:2003tba}. One of the way to explain the observed acceleration of the Universe is to develop alternative (modified) theories of gravity which can be reduced to GR on some scales but can lead to Universe acceleration on very large scales. For recent reviews of modified theories of gravity, see, for instance,\cite{Nojiri:2010wj}, \cite{Nojiri:2017ncd}, \cite{Joyce:2014kja} and the fundamental work \cite{Clifton:2011jh}.

The growing interest to modified theory of gravity is due to detection of gravitational waves are emitted  by the system of two black holes (BHs) which allows to test extensions of the GR theory \cite{Abbott:2016blz}. Therefore, besides of cosmological applications of modified gravity theories, it is of interest to study spherically symmetric solutions as possible resulting BH which generates gravitational waves, may be with other characteristics then in GR. In our present investigation we analyze spherically symmetric solutions in the theory of gravity with high-order corrections to scalar curvature.

 Modifications of GR that include higher-order curvature corrections to the
 Einstein-Hilbert action have a natural explanation, related to taking into account quantum effects
 in the low-energy limit of string theory, superstrings, and supergravity, needed for the construction
 of a quantum theory of gravity \cite{cfo-Baumann:2014nda}.

 We can mention that a prominent example
 of application of quantum corrections to GR, with derivatives up to fourth order,
 has been demonstrated
  by A. Starobinsky \cite{cfo-Starobinsky:1980te} in cosmology. It was shown that such corrections may control an accelerated expansion of the Universe at its early stage of evolution (inflation).

  This kind of models has been developed towards considering 6th-order corrections in theories of gravity
  of the kind  $R+\alpha R^2  +\gamma R\Box R$, where $\alpha$ and $\gamma$ are some constants,
  and the additional terms that modify the Einstein theory were, with the aid of conformal transformations
 of the metric, put into correspondence to certain effective scalar fields \cite{cfo-Gottlober:1989ww}--\cite{cfo-Cuzinatto:2013pva},
 which led to a two-field treatment of such models. Also, in \cite{cfo-Castellanos:2018dub}
 the correction  $R\Box R$ was treated as a small perturbation, and its influence on the parameters
 of cosmological perturbations was studied.

 Inclusion of higher derivatives to modified theories of gravity is also caused by the necessity of introducing
 such corrections at renormalization of the stress-energy tensors of quantum fields in the framework
 of the semiclassical approach to gravity \cite{cfo-BOS1992}.

In the article \cite{cfo-Naruko:2015zze} it was shown the way how to reduce theory of gravity which contain in the action the Ricci scalar and its first and second derivatives to GR minimally coupled to few scalar fields. For special choices of functional dependence $f(R,(\nabla R)^2,\Box R)$ it is possible to reduce the theory to chiral cosmological model. One such example and its application in cosmology is demonstrated in \cite{cf-CHNM2017},\cite{cfo-Chervon:2018ths},\cite{Chervon:2019jfu}.

In the work \cite{Chervon:2019jfu} the study of the model, using the technique described in \cite{cfo-Naruko:2015zze}, \cite{cfo-Tsoukalas2017},
 was continue. There were carried out a similar transition to a scalar-tensor theory by introducing Lagrangian multipliers and
 the arising auxiliary fields. Using a conformal transformation from the Jordan frame
 to the Einstein one there were obtained that the model can be represented as a two-component nonlinear
 sigma model with an interaction potential. Such model is named as a chiral cosmological model (CCM) for consideration in cosmology.
 Because in our present work we deal with gravitation in spherically symmetric spacetime, not in cosmology, the model  with the action
 \begin{equation}\label{act-csgm}
  S=\int\sqrt{-g}d^4x\left(\frac{R}{2\kappa}-
    \frac{1}{2}h_{AB}(\varphi)\varphi^A_{,\mu}\varphi^B_{,\nu}g^{\mu\nu}-W(\varphi)\right),
\end{equation}
 we will call
 {\it Chiral Self-Gravitating Model} (CSGM). It is clear that CCM is the case of CSGM in cosmological spaces.
 Thus, the main goal of our investigation is to find conditions to have exact static, spherically symmetric solutions in the CSGM.

The article is organised as follow. In Sec. 2 we represent the action of the $f(R)$ gravity with kinetic curvature scalar in Einstein's frame and give connection of the model to CSGM. The Sec. 2 is devoted to  tensor equations of CSGM in metric spacetime which represented in Sec. 3 for spherically symmetric spacetime in harmonic coordinates. Sec. 4 contains the gravitation and fields equation in spherically symmetric spacetime in harmonic coordinate. In Sec. 5 we consider the case of scaling transformation when the model transforms from Jordan to Einstein frame and the representative function $f_1$ of the scalar field $\phi$ is chosen as GR analog, namely $f_1(\phi)=\phi/2$. For this case it was found the three classes of solutions for gravitational field and demonstrated the method of determination of kinetic function $X(\phi)$ corresponding to known scalar field dependence on analog to radial coordinate $u$. Sec. 6 devoted to investigation of the model subject to special relation (ansatz) between metric components. Three classes of solutions are represented as well. In Conclusion we discuss obtained results and future investigations.

\vspace{2mm}

\section{The model in Einstein frame}

  In the work \cite{cfo-Naruko:2015zze} the authors consider the most general form of $f(R)$ gravity with higher derivatives of first and second order with respect to Ricci scalar.
 Such a theory of gravity is described by the action
\beq           \label{cfo-Gen}
	S_{\rm gen} = \int d^{4}x\sqrt{-g}\, f \left(R,(\nabla R)^{2},\square R \right),
\eeq
where $(\nabla R)^{2}=g^{\mu\nu}\nabla_{\mu}R\nabla_{\nu}R$ and $\square R=\nabla_{\mu}\nabla^{\mu} R$.

The truncated model with the action
\beq           \label{cfo-bfR}
	S_{fRR'}=\int d^{4}x\sqrt{-g}\, f\left(R,(\nabla R)^{2} \right),
\eeq
 where
\beq\label{cfo-f1X}
f(R,(\nabla R)^2)=f_1(R)+X(R)\nabla_\mu R\nabla^\mu R.
\eeq
have been studied for cosmological applications in \cite{cf-CHNM2017}, \cite{cfo-Chervon:2018ths}, \cite{Chervon:2019jfu}. In \eqref{cfo-f1X} and hereafter $f_1(R)$ and $X(R)$ are $\textsl{C}^1$ functions of scalar curvature.

The model \eqref{cfo-bfR}-\eqref{cfo-f1X}
 can be transformed to Einstein scalar fields gravity \cite{cfo-Naruko:2015zze},\cite{Chervon:2019jfu} with the action
\begin{eqnarray}\nonumber
 S_{EfRR'}=\int d^4x \sqrt{-g_E}\bigg(\dfrac{R_E}{2} - \dfrac{1}{2}g^{\mu\nu}_E\chi_{,\mu}\chi_{,\nu}
 + \dfrac{1}{4}f_1(\phi)e^{-2\sqrt{2/3}\chi}-
 \dfrac{1}{4}\phi e^{-\sqrt{2/3}\chi}\\
 \label{act-2}
  + \dfrac{1}{2}Xe^{-\sqrt{2/3}\chi} g^{\mu\nu}_E\phi_{,\mu}\phi_{,\nu} \bigg).
\end{eqnarray}
where subscript $"E"$ denotes the Einstein's frame. In \eqref{act-2} a new scalar field have been introduced by the relation $\lambda = \exp (\sqrt {2/3} \chi)$, to obtain
  a canonical form for the scalar field $\chi$ in the action with $\lambda$-field. Non-minimal coupling to gravity via scalar field $\lambda$ connected with conformal transformation $g^{E}_{\mu\nu}=\Omega^2g^{J}_{\mu\nu} $ as $\Omega^2=2\lambda $. The second scalar field $\phi$ takes the nonlinearity of $f_1(R)$ and in some sense can be in correspondence with dependence $f_1(\phi)$ \cite{cfo-Naruko:2015zze},\cite{Chervon:2019jfu}.

The action $S_{EfRR'}$ (\ref{act-2}) can be presented in the form of a chiral cosmological model \cite{cfo-chervon2013} with two chiral fields
 $\varphi^1=\chi$, $\varphi^2=\phi$, and 2D metric of the target space
\begin{equation}
	ds^2_{\sigma} = d\chi^2 - e^{-\sqrt{2/3}\chi}X(\phi )d \phi^2.
\end{equation}
Also it needs to include in the model the interaction potential
\begin{equation} W=\dfrac{1}{4}e^{-\sqrt{2/3}\chi}\left(\phi-e^{-\sqrt{2/3}\chi}f_1(\phi)\right).
\end{equation}

Because in our present work we deal with gravitation in spherically symmetric spacetime, not in cosmology, the model  (\ref{act-2}) and similar ones are nothing but self-gravitating non-linear sigma model with the potential of interaction. For the sake of brevity we will refers to this type models as {\it Chiral Self-Gravitating Model} (CSGM). It is clear that CCM is the case of CSGM in cosmological spaces.

In the next section we will represent basic equation of CSGM.

\section{Chiral Self-Gravitating Model}\label{CCM}

The action of CSGM as the action of the self-gravitating nonlinear sigma model with the
potential of (self)interactions $W(\varphi)$ \cite{Chervon95b}, \cite{Chervon95a}
reads:
\begin{equation}\label{act-ccm}
  S_{CSGM}=\int\sqrt{-g}d^4x\left(\frac{R}{2\kappa}-
    \frac{1}{2}h_{AB}(\varphi)\varphi^A_{,\mu}\varphi^B_{,\nu}g^{\mu\nu}-W(\varphi)\right),
\end{equation}
where $ R $ being a scalar curvature of a Riemannian manifold with the metric
$g_{\mu\nu}(x)$, $\kappa$ -- Einstein gravitational constant, $\varphi=(\varphi^1,\ldots,\varphi^N)$~being a
multiplett of the chiral fields (we use a notation
$\varphi^A_{,\mu}=\partial_{\mu}\varphi^A =\frac{\partial
  \varphi^A}{\partial x^\mu}$),
$h_{AB}$~being the metric of the target space (the chiral space) with
the line element
\beq\label{tsmet}
 d s^2_{\sigma}=h_{AB}(\varphi)d\varphi^A d\varphi^B,~~A,B,\ldots =\overline{1,N}.
\eeq
The energy-momentum tensor for the model (\ref{act-ccm}) reads
\begin{equation}\label{ch-em}
  T_{\mu\nu}=h_{AB}\varphi^{A}_{,\mu}\varphi^B_{,\nu}-
  g_{\mu\nu}\left(\frac{1}{2}\varphi^A_{,\alpha}\varphi^B_{,\beta}g^{\alpha\beta}h_{AB}+
    W(\varphi)\right).
\end{equation}

The Einstein equation can be represented in the form
\beq\label{ein}
R_{\mu\nu}=\varkappa\{
h_{AB}\varphi^A_{,\mu}\varphi^B_{,\nu}+ g_{\mu \nu} W(\varphi)\}
\eeq
which simplify the derivation of gravitational dynamic equations.

Varying the action (\ref{act-ccm}) with respect to $\varphi^C$, one can
derive the dynamic equations of the chiral fields
\begin{equation}\label{4}
  \frac{1}{\sqrt{-g}}\partial_{\mu}(\sqrt{-g}h_{AB}g^{\mu\nu}\varphi_{,\nu}^A)-\frac{1}{2}\frac{\partial
    h_{BC}}{\partial\varphi^A}\varphi^C_{,\mu}\varphi^{B}_{,\nu}g^{\mu\nu}-W_{,A}=0,
\end{equation}
where $W_{,A}=\frac{\partial W}{\partial\varphi^A}$.
Considering the action (\ref{act-ccm}) in the framework of
cosmological spaces, we arrive to a chiral cosmological model \cite{cfo-chervon2013},
\cite{Chervon95b}, \cite{Chervon95a}, \cite{ch00mg}, \cite{ch02gc},\cite{cfo-CHFK2015}.
%

\vspace{2mm}
\section{CSGM in spherically symmetric spacetime}

The standard representation of the spherically symmetric spacetime without gauge specification is \cite{bronnikov08}
\begin{equation}\label{SphSy_mbr}
  ds^2=-e^{2\nu (u)}dt^2+e^{2\lambda (u)} du^2 +e^{2\beta (u)} \left(d\theta^2+\sin^2 \theta d\varphi^2 \right)
\end{equation}

Let us choose the harmonic coordinates where we have the connection between metric function of the form
\begin{equation}
\lambda=2\beta+\nu
\end{equation}

Einstein equations \eqref{ein} can be derived based on components of Ricci tensor and they are:
\begin{equation}\label{Eh-0br}
e^{-4\beta}\nu'' =-\varkappa e^{2\nu}\dfrac{1}{4}e^{-\sqrt{2/3}\chi}\left(\phi-e^{-\sqrt{2/3}\chi}f_1(\phi)\right),
\end{equation}

\begin{equation}\label{Eh-1br}\nonumber
-2\beta''-\nu''+2(\beta')^2 +4\beta'\nu'=
\end{equation}
\begin{equation}
=\varkappa
\left(h_{11}(\chi')^2- e^{-\sqrt{2/3}\chi}X(\phi)(\phi')^2+e^{4\beta+2\nu} \dfrac{1}{4}e^{-\sqrt{2/3}\chi}\left(\phi-e^{-\sqrt{2/3}\chi}f_1(\phi)\right)\right),
\end{equation}

\begin{equation}\label{Eh-2br}
1- \beta ''\exp \left[-2 \beta-2\nu\right]   =\varkappa e^{2\beta}\dfrac{1}{4}e^{-\sqrt{2/3}\chi}\left(\phi-e^{-\sqrt{2/3}\chi}f_1(\phi)\right).
\end{equation}

Chiral fields equations after substitution of metric components and the potential yield
\begin{equation}
h_{11}\chi'' -\frac{1}{\sqrt{6}} e^{-\sqrt{\frac{2}{3}}\chi}X(\phi)(\phi')^2
-e^{4\beta+2\nu}\left[-\frac{1}{2\sqrt{6}} e^{-\sqrt{\frac{2}{3}}\chi}\phi+\frac{1}{\sqrt{6}} e^{-2\sqrt{\frac{2}{3}}\chi}f_1(\phi) \right]=0,
\end{equation}
\begin{equation}
-\sqrt{\frac{2}{3}}X(\phi)\phi'\chi'+\frac{1}{2}X_\phi'(\phi')^2+
X(\phi)\phi''
+\frac{1}{4}e^{4\beta+2\nu}\left(1-e^{-\sqrt{\frac{2}{3}}\chi}f_{1,\phi}\right)=0.
\end{equation}

From equations \eqref{Eh-0br} and \eqref{Eh-2br} one can obtain the relation which make restriction on the metric components
\begin{equation}\label{b-nu}
\beta''-\nu''=e^{2\beta+2\nu}.
\end{equation}

\section{The case of scaling transformation}
Following by special suggestion in cosmology \cite{cfo-Tsoukalas2017}, \cite{Chervon:2019jfu}  let us study the case when the scalar field  $\chi$ is equal to special constant $\chi= -\sqrt{\frac{3}{2}}\ln 2$. This value of $\chi$ corresponded to identical conformal transformation with $\Omega^2=1$.

The gravitational and chiral field equations will take the following form
\begin{equation}\label{30-nu}
e^{-(4\beta+2\nu)}\nu''=-\varkappa \left(\phi/2-f_1(\phi)\right),
\end{equation}

\begin{equation}\label{31-X}
-2\beta''-\nu''+2(\beta')^2 +4\beta'\nu'=\varkappa \left(-2X(\phi)(\phi')^2+e^{4\beta+2\nu}(\phi/2-f_1(\phi))\right),
\end{equation}

\begin{equation}\label{32-beta}
1-\beta'' e^{-2\beta-2\nu}=\varkappa e^{2\beta}(\phi/2-f_1(\phi)),
\end{equation}

\begin{equation}\label{f-2-spec}
-2X(\phi) (\phi')^2-e^{4\beta+2\nu}(-\phi+4f_1(\phi))=0,
\end{equation}

\begin{equation}\label{34-phi}
2X(\phi)\phi''+X'_\phi (\phi')^2+e^{4\beta+2\nu}(1/2-f_{1,\phi})=0.
\end{equation}

The equation \eqref{f-2-spec} may be considered as additional one, for the model with one field $\phi$ this equation will not appear (no variation on $\chi$).

\subsection{Case $f_1(\phi)=\phi/2$}

Such suggestion bring the model close to GR, if we additionally set $X(\phi)=0$.
For the case $f_1(\phi)=\phi/2$ we have the system of equations
\begin{equation}\label{nu-1}
\nu''=0,~~\nu=A_1u+A_2,~~A_1, A_2 - const.,
\end{equation}

\begin{equation}\label{35}
-2\beta''+2(\beta')^2 +4\beta'A_1=\varkappa \left(-2X(\phi)(\phi')^2\right),
\end{equation}

\begin{equation}\label{sol-beta}
1-\beta'' e^{-2\beta-2A_1u-2A_2}=0,
\end{equation}

\begin{equation}\label{37}
2X(\phi)\phi''+X'_\phi (\phi')^2=0.
\end{equation}

\subsubsection{The solutions of the auxiliary equation}\label{aux-eq}
Let us study solutions of the equation
\begin{equation}\label{y-abu}
y''(u)=a e^{2by},~~ a,b -const.,
\end{equation}
which we will use for special values of parameters later on.

The first integral of \eqref{y-abu} is
\begin{equation}\label{sol-beta-31}
(y')^2=\frac{a}{b} e^{2by}+C_1.
\end{equation}

There are three possibilities according with $C_1$ value.

\begin{itemize}
\item $C_1=0$

The solution is
\begin{equation}\label{y-abu-1}
y(u)=-\frac{1}{b}\ln \left(\pm \sqrt{ab}(u-u_*)\right),
\end{equation}
where the integration constant $u_*$ may change the sign of the argument $(u-u_*)$.

\item $C_1=\mu^2>0$

The solution for this case is
\begin{equation}\label{y-abu-2}
y(u)=\mp \mu u-\frac{1}{b}\ln \left( -\frac{a}{b} +\frac{1}{4\mu^2} e^{2b\mu u}\right).
\end{equation}
From here we set $u=u-u_*$ to simplify the expressions.
\item $C_1=-\alpha^2<0 $

The solution is
\begin{equation}\label{y-abu-3}
y(u)=-\frac{1}{b}\ln \left(\sqrt{\frac{b}{a}}\frac{\alpha}{\cos (\alpha b u)} \right).
\end{equation}
\end{itemize}


\subsubsection{Solutions of the model}
The equation \eqref{sol-beta} can be reduced to
\begin{equation}\label{sol-beta-2}
y''= 2e^{y}
\end{equation}
by the substitution
\begin{equation}\label{sol-beta-3}
y=2\beta+2A_1u+2A_2.
\end{equation}

Thus we have three solutions for $y(u)$ by setting in the solutions above $a=2,~b=1/2$.
\begin{itemize}
\item $C_1=0$

The solution is
\begin{equation}\label{y-12u-1}
y(u)=-2\ln \left(u \right).
\end{equation}

From here one can find
\begin{equation}\label{be-12u-1}
\beta(u)=-\ln (u)-A_1 u-A_2.
\end{equation}

Equation \eqref{35} gives
\begin{equation}\label{A1-X-1}
\varkappa X(\phi)(\phi')^2=A_1^2.
\end{equation}

Thus we can represent class of solutions by giving $\phi=\phi (u)$.

In literature \cite{bronnikov08} one can find the following type of solutions for the scalar field
\begin{equation}\label{phi1-2-u}
\phi_1 (u)= \alpha_* u +\phi_*,~~\phi_2(u) = \beta_* \tanh (\lambda_* u)
\end{equation}
and
\begin{equation}\label{phi3-u}
\phi_3 (u)=D \arctan \frac{\sqrt{u^2-p^2}}{p}+\phi_*,~~D, p =const.
\end{equation}
Here the letters with lower star mean constant. Note that the function $f_1(\phi)$ can be easy restored for given $\phi$ as
$$
f_1(\phi)= \frac{1}{2}\phi .
$$
Note that the potential for this case is equal to zero and we deal with massless scalar field.

It is easy to define the function $X(\phi)$ for each case.
The sign of kinetic function $X(\phi)$ corresponds to type of the scalar field: canonical, if $X<0$, and phantom, if $X>0$. Thus we have corresponding to the scalar fields kinetic functions:
\begin{equation}\label{0X-phi1}
X_1(\phi_1)=\frac{A_1^2}{\varkappa \alpha_*^2}=const.
\end{equation}

\begin{equation}\label{0X-phi2}
X_2(\phi_2)  = \frac{A_1^2\beta_*^2}{\varkappa \lambda_*^2}\left(\beta_*^2-\phi_2^2(u)\right)^{-2}.
\end{equation}

\begin{equation}\label{0X-phi3}
X_3(\phi_3)=\frac{A_1^2p^2}{\varkappa D^2}\tan^2\left(\frac{\phi_3(u)-\phi_*}{D}\right) \left(1+\tan^2\left(\frac{\phi_3(u)-\phi_*}{D}\right)\right).
\end{equation}

\item $C_1=\mu^2>0$

The solution for this case is
\begin{equation}\label{y-12u-2}
y(u)=\mp \mu u-\ln \left( -2 +\frac{1}{4\mu^2} e^{2\mu u}\right).
\end{equation}
From here one can find
\begin{equation}\label{be-12u-2}
\beta(u)=\frac{1}{2}\left[\mp \mu u-\ln \left( -2 +\frac{1}{4\mu^2} e^{2\mu u}\right)\right]-A_1 u-A_2.
\end{equation}

Equation \eqref{35} gives
\begin{equation}\label{A1-X-2}
\varkappa X(\phi)(\phi')^2=-\left(\frac{\mu^2}{4}-A_1^2\right)
\end{equation}

The solutions for kinetic function $X(\phi)$ are
\begin{equation}\label{+X-phi1}
X_1(\phi_1)=\frac{\left(A_1^2-\frac{\mu^2}{4}\right)}{\varkappa \alpha_*^2}=const.
\end{equation}

\begin{equation}\label{+X-phi2}
X_2(\phi_2)  = \frac{\left(A_1^2-\frac{\mu^2}{4}\right)\beta_*^2}{\varkappa \lambda_*^2}\left(\beta_*^2-\phi_2^2(u)\right)^{-2}.
\end{equation}

\begin{equation}\label{+X-phi3}
X_3(\phi_3)=\frac{\left(A_1^2-\frac{\mu^2}{4}\right)p^2}{\varkappa D^2}\tan^2\left(\frac{\phi_3(u)-\phi_*}{D}\right) \left(1+\tan^2\left(\frac{\phi_3(u)-\phi_*}{D}\right)\right).
\end{equation}

\item $C_1=-\alpha^2<0 $

The solution is
\begin{equation}\label{y-12u-3}
y(u)=2\ln \left(\frac{1}{2}\frac{\alpha}{\cos \left(\frac{\alpha u}{2}\right)} \right).
\end{equation}

From here one can find
\begin{equation}\label{be-12u-3}
\beta(u)=\ln \left(\frac{1}{2}\frac{\alpha}{\cos \left(\frac{\alpha u}{2}\right)} \right)-A_1 u-A_2.
\end{equation}

Equation \eqref{35} gives
\begin{equation}\label{A1-X-3}
\varkappa X(\phi)(\phi')^2=\frac{\alpha^2}{4}+A_1^2.
\end{equation}

The solutions for kinetic function $X(\phi)$ are
\begin{equation}\label{-X-phi1}
X_1(\phi_1)=\frac{\left(A_1^2+\frac{\alpha^2}{4}\right)}{\varkappa \alpha_*^2}=const.
\end{equation}

\begin{equation}\label{-X-phi2}
X_2(\phi_2)  = \frac{\left(A_1^2+\frac{\alpha^2}{4}\right)\beta_*^2}{\varkappa \lambda_*^2}\left(\beta_*^2-\phi_2^2(u)\right)^{-2},
\end{equation}

\begin{equation}\label{-X-phi3}
X_3(\phi_3)=\frac{\left(A_1^2+\frac{\alpha^2}{4}\right)p^2}{\varkappa D^2}\tan^2\left(\frac{\phi_3(u)-\phi_*}{D}\right) \left(1+\tan^2\left(\frac{\phi_3(u)-\phi_*}{D}\right)\right).
\end{equation}
\end{itemize}

It should be stressed that equation \eqref{37} is satisfied when $X(\phi)(\phi')^2=const.$
Indeed, taking the derivative with respect to $u$ from both part of equations \eqref{A1-X-1},\eqref{A1-X-2} and \eqref{A1-X-3} one can find $X'(\phi')^3=-2X\phi'\phi''$
and equation \eqref{37} is automatically satisfied.

Thus we can state that $X(\phi)(\phi')^2=const.$ with \eqref{nu-1}, \eqref{be-12u-1}, \eqref{be-12u-2} and \eqref{be-12u-3} give us three solutions. Therefore for each given dependence of the scalar field $\phi$ as function on $u$ one can define the kinetic function $X(\phi)$ which holds the solution. Examples of such solutions we described above.

\section{Ansatz $\beta=m \nu $}

Choosing the relation between metric functions
\begin{equation}\label{anz1}
\beta = m \nu,~~ m=const.
\end{equation}
we can solve the equation \eqref{b-nu} which takes the following form
\begin{equation}\label{dd-nu}
\nu''=\frac{1}{m-1}e^{2\nu (m+1)}.
\end{equation}

So we can use the solutions of auxiliary equation \eqref{y-abu} with $y=\nu $ and
$$
a=\frac{1}{m-1},~~b=m+1,~~m\ne 1,~~m\ne -1.
$$

With the ansatz \eqref{anz1} and choosing the $f_1(\phi)$ as
\begin{equation}\label{f2-def}
f_1(\phi)=\phi/2+K_2 f_2(\phi),~~ K_2=const.,
\end{equation}
the gravitational and field equations take the following form
\begin{equation}\label{30-nu-az}
e^{-2(2m+1)\nu}\nu''=\varkappa K_2 f_2(\phi),
\end{equation}

\begin{equation}\label{31-X-az}
-(2m+1)\nu''+2m(m+2)(\nu')^2=\varkappa \left(-2X(\phi)(\phi')^2-e^{2(2m+1)\nu} K_2 f_2(\phi)\right),
\end{equation}

\begin{equation}\label{32-beta-az}
1-m\nu'' e^{-2(m+1)\nu}=-\varkappa e^{2m\nu}K_2 f_2(\phi),
\end{equation}

\begin{equation}\label{34-phi-az}
2X(\phi)\phi''+X'_\phi (\phi')^2-e^{2(2m+1)\nu}K_2f_{2,\phi}=0.
\end{equation}

Taking into account \eqref{dd-nu} equation \eqref{30-nu-az} leads to
\begin{equation}\label{f-2-nu}
f_2(\phi)=\frac{1}{\varkappa K_2 (m-1)}e^{-2m\nu}.
\end{equation}

Substitution \eqref{f-2-nu} into \eqref{32-beta-az} gives the identity.
Also  substitution \eqref{f-2-nu} into \eqref{31-X-az} leads to
\begin{equation}\label{31-X-azm}
-(2m+1)\nu''+2m(m+2)(\nu')^2 +\frac{1}{m-1}e^{2(m+1)\nu}=-2\varkappa X(\phi)(\phi')^2.
\end{equation}

Thus we can find the dependence $f_2=f_2(\phi(u))=f_2(u)$ from \eqref{f-2-nu} using known the solutions of equation \eqref{dd-nu} on the basis of auxiliary equation solutions \eqref{aux-eq}.

We can obtain the general form for the combination $X(\phi)(\phi')^2$ if we insert $\nu''$ from \eqref{dd-nu}
and using the first integral of \eqref{dd-nu}
\begin{equation}\label{nu-m1}
\nu'^2=\frac{1}{m^2-1} e^{2(m+1)\nu}+C_1
\end{equation}
into equation \eqref{31-X-az}.
The result is

\begin{equation}\label{Xphi}
\left[\frac{2m}{m^2-1}\right] e^{2(m+1)\nu}  + 2m(m+2)C_1=-2\varkappa X(\phi)(\phi')^2.
\end{equation}

Let us note that
we can not obtain the constant combination $X(\phi)(\phi')^2=const.$  Thus the case considered in previous section  is not compatible with ansatz approach.

\subsection{The solution $C_1=0$}
We may consider two branches of solution in this case. Namely
\begin{equation}\label{C1=0a}
\nu_1(u)= -\frac{1}{m+1}\ln \left(-\sqrt{\frac{m+1}{m-1}}(u-u_*)\right),~~u<u_*,
\end{equation}

\begin{equation}\label{C1=0b}
\nu_2(u)= -\frac{1}{m+1}\ln \left(\sqrt{\frac{m+1}{m-1}}(u-u_*)\right),~~u>u_*,
\end{equation}
where $u_*$ is an integration constant.

From \eqref{30-nu} we obtain
\begin{equation}\label{f2K-0}
f_2(\phi)=\frac{1}{\varkappa K_2 (m-1)}e^{-2m\nu}.
\end{equation}
Substituting the solution \eqref{C1=0a} and  \eqref{C1=0b} in \eqref{f2K-0} one can find

\begin{equation}\label{f2K-01}
f_2(\phi)=\left(\varkappa K_2 (m-1)\right)^{-1}
\left[ \left(\frac{m+1}{m-1}\right)(u-u_*)^2\right]^{\frac{m}{m+1}}.
\end{equation}

Substituting the solution \eqref{C1=0a} and  \eqref{C1=0b} in \eqref{Xphi} one can find

\begin{equation}
X(\phi)(\phi')^2=-\frac{2m}{2\varkappa (m+1)^2 (u-u_*)^2}. 
\end{equation}

Ones again, let us define the kinetic function for scalar field \eqref{phi1-2-u} and \eqref{phi3-u}.
\begin{equation}\label{0Xm-phi1}
X(\phi_1)=-\frac{2m}{2\varkappa (m+1)^2 (\phi_1(u)-\phi_*)^2},
\end{equation}

\begin{equation}\label{0Xm-phi2}
X(\phi_2)=-\frac{2m\beta_*^2}{2\varkappa (m+1)^2 (\beta_*^2-\phi^2_2)^2 (\tanh^{-1}(\phi_2/\beta_*)^2 }.
\end{equation}
Here $\tanh^{-1}x:= {\rm arctanh}\,x$.

\begin{equation}\label{0Xm-phi3}
X_3(\phi_3)=-\frac{2m\tan^2 ((\phi_3-\phi_*)/D)}{2\varkappa (m+1)^2 D^2 p^2}.
\end{equation}
Corresponding solutions for $f_2(\phi)$ are
\begin{equation}\label{f2-1-m}
f_2(\phi_1)=\left(\varkappa K_2 (m-1)\right)^{-1}
\left[ \frac{m+1}{m-1}\right]^{\frac{m}{m+1}}\left(\frac{\phi_1(u)-\phi_*}{\alpha_*}
\right)^{\frac{2m}{m+1}},
\end{equation}

\begin{equation}\label{f2-2-m}
f_2(\phi_2)=\left(\varkappa K_2 (m-1)\right)^{-1}
\left[ \frac{m+1}{m-1}\right]^{\frac{m}{m+1}}\left(\frac{1}{\lambda_*}
\tanh^{-1}\left(\frac{\phi_2(u)}{\beta_*}\right)\right)^{\frac{2m}{m+1}},
\end{equation}

\begin{equation}\label{f2-3-m}
f_2(\phi_3)=\left(\varkappa K_2 (m-1)\right)^{-1}
\left[ \frac{m+1}{m-1}\right]^{\frac{m}{m+1}}\left[\frac{1}{p}
\cos \left(\phi_3 (u)-\phi_*\right)\right]^{-\frac{2m}{m+1}}.
\end{equation}

From the presentation \eqref{f2-def} it is clear that the function $f_2(\phi)$ gives, in some sense, the deviation from GR. Thus we can see from \eqref{f2-1-m}-\eqref{f2-3-m} various functional possible addition, keeping in mind exchange $\phi \rightarrow R$.

\subsection{The solution $C_1=\mu^2>0$}\label{sbs6.1}
The solution
\begin{equation}\label{u-C1=mu}
u-u_*=\mp \frac{1}{(m+1)|\mu|}\ln \left[ \frac{2\mu^2+2|\mu|\sqrt{\frac{z^2}{m^{2}-1}+\mu^2}}{z}\right],~~
z=e^{(m+1)\nu},
\end{equation}
must be inversed in order to find the
solution for $\nu (u)$. From \eqref{u-C1=mu} one can obtain

\begin{equation}
\nu=\mu u -\frac{1}{m+1}\ln \left( \frac{m+1}{m-1}+\frac{1}{\mu^2}e^{2\mu(m+1)u}\right).
\end{equation}

Then from \eqref{Xphi} we find

\begin{equation}\label{Xphi-1}
2\varkappa X(\phi)(\phi')^2= -\left[\frac{2m}{m^2-1}\right] \frac{\mu^2 E_1(u)}{\left(\frac{\mu^2}{m^2-1}+E_1(u)\right)^2}-
2\mu^2 m(m+2),
\end{equation}
where
$$
E_1(u)=e^{2\mu(m+1)u}.
$$

Thus we can state that for each given $\phi(u)\ne const.$ there are exist  $X(\phi)$, obtained from  \eqref{Xphi-1}  and which gives the exact solution with
\begin{equation}
f_2(\phi (u))=\frac{1}{\varkappa K_2 (m-1)}e^{-2m\mu u}\left( \frac{1}{m^2-1}+\frac{1}{\mu^2}E_1(u)\right)^{\frac{2m}{m+1}}.
\end{equation}

Inverting the dependence $\phi $ on $u$ as $u(\phi)$ the dependence $f_2$ on $\phi$ will be restored. The kinetic function $X(\phi)$ can be defined by algorithm described in Sec. \ref{sbs6.1}

\subsection{The solution $C_1=-\alpha^2<0$}
The solution
\begin{equation}\label{u-C1-al}
u-u_*=\mp \frac{1}{\alpha(m+1)} \arctan\left(\frac{1}{\sqrt{\frac{z^{2}}{m^{2}-1}+\alpha^2}}\right),~~z=e^{(m+1)\nu},
\end{equation}
must be inversed in order to find the
solution for $\nu (u)$. From \eqref{u-C1-al} one can obtain

\begin{equation}
\nu (u)=\frac{1}{m+1}\ln \left[\frac{(m+1)\alpha^2}{(m-1)}\left(1 \pm \tan^2(\alpha(m+1)u)\right)\right].
\end{equation}

We can obtain$f_2(\phi(u))$ from \eqref{f-2-nu}

\begin{equation}
f_2(\phi (u))=\left[\alpha^2 (m-1)\left(1 \pm \tan^2(\alpha (m+1) u)\right)\right]^{-\frac{2m}{m+1}}\left(\varkappa K_2 (m-1)\right)^{-1}.
\end{equation}

From \eqref{Xphi} one can find
\begin{equation}
2\varkappa X(\phi)(\phi')^2=-\alpha^4 2m(m^2-1)\left[ 1 \pm \tan^2(\alpha (m+1) u)\right]^2-2\alpha^2 m(m+2).
\end{equation}

Inverting the dependence $\phi $ on $u$ as $u(\phi)$ the dependence $f_2$ on $\phi$ will be restored. The kinetic function $X(\phi)$ can be defined by algorithm described in Sec. \ref{sbs6.1}.

\section{Conclusion}

In this paper we presented an analysis of non-linear self-gravitating sigma model with the potential of interaction (chiral self-gravitating model) of $f(R)$ gravity with a kinetic curvature scalar in a spherically symmetric spacetime. We derived the model's equation in harmonic coordinates and found restriction on the metric components for this case. For the case of scaling transformation, when Weil conformal function is equal to one, it was found metric components for the case when $f_1(\phi)=\phi/2$ and pointed out the method of definition the kinetic function $X(\phi)$ if the dependence of scalar field $\phi$ on radial-like coordinate $u$ is known. New solutions for the metric components and explanation of kinetic function $X(\phi)$ derivation were founded for special relation between metric components (ansatz) $\beta = m \nu,~ m=const.$ Further we plan to consider geometrical structure of obtained solutions such as horizons and singularities.
\vspace{5mm}

\noindent{\bf Acknowledgments}

S.V.C. is thankful to CNPq for support his visit to Brazil, Vitoria, UFES, where the work was finalised. J.C.F. thanks CNPq (Brazil) and FAPES (Brazil).
  S.V.C. and F.I.V. are thankful for the financial support of RFBR by grant 20-02-00280-A. S.V.C. is grateful for the support of the Program of Competitive Growth of Kazan Federal University.

\bigskip


\begin {thebibliography}{900}	

\small

\bibitem{cfo-Perlmutter:1998np}
	Perlmutter S. et al. [Supernova Cosmology Project Collaboration].
	Measurements of Omega and Lambda from 42 high redshift supernovae.
	{\it Astrophys. J.}, 1999, vol. 517, pp. 565--586.

\bibitem{cfo-Riess:1998cb}
	Riess A.G. et al. [Supernova Search Team].
Observational evidence from supernovae for an accelerating universe and a cosmological constant.
{\it Astron. J.}, 1998, vol. 116, pp. 1009--1038.


\bibitem{Ade:2015xua}
Ade P.A.R.  et al. [Planck collaboration].
Planck 2015 results. XIII. Cosmological parameters,
{\it Astron. Astrophys.}, 2016, vol. 594, A13.

\bibitem{BICEP2-2015}
Ade P.A.R. et al. [BICEP2 and Keck Array collaborations].
 Improved constraints on cosmology and foregrounds from BICEP2 and Keck Array cosmic microwave background data with inclusion of 95GHz band.
 {\it Phys. Rev. Lett.}, 2016, vol. 116, p. 031302.

\bibitem{WMAP2013}
Hinshaw G. et al. [WMAP collaboration].
Nine-year Wilkinson Microwave Anisotropy Probe (WMAP) observations: cosmological parameter results.
{\it Astrophys. J. Suppl.}, 2013, vol. 208, p. 19.

\bibitem{Seljak:2004xh}
Seljak U. et al. [SDSS collaboration].
Cosmological parameter analysis including SDSS Ly-alpha forest and galaxy bias: constraints on the primordial spectrum of fluctuations, neutrino mass and dark energy.
{\it Phys. Rev. D}, 2005, vol. 71, p. 103515.

\bibitem{Eisenstein:2005su}
Eisenstein, D.J. et al. [SDSS collaboration].
Detection of the baryon acoustic peak in the large-scale correlation function of SDSS luminous red galaxies.
{\it Astrophys. J.}, 2005, vol. 633, pp. 560--574.

\bibitem{Jain:2003tba}
Jain B. and Taylor A.
Cross-correlation tomography: measuring dark energy evolution with weak lensing.
{\it Phys. Rev. Lett.}, 2003, vol. 91, p. 141302.

\bibitem{Nojiri:2010wj}
Nojiri S. and Odintsov S.D.
Unified cosmic history in modified gravity: from F(R) theory to Lorentz non-invariant models.
{\it Phys. Rept.}, 2011, vol. 505, pp. 59--144.

\bibitem{Nojiri:2017ncd}
Nojiri S. and Odintsov S.D. and Oikonomou V.K.
Modified gravity theories on a nutshell: inflation, bounce and late-time evolution.
{\it Phys. Rept.}, 2017, vol. 692, pp. 1--104.

\bibitem{Joyce:2014kja}
Joyce A., Jain B., Khoury J. and Trodden M.
Beyond the cosmological Standard Model.
{\it Phys. Rept.}, 2015, vol. 568, pp. 1--98.

\bibitem{Clifton:2011jh}
Clifton T., Ferreira, P. G., Padilla A. and Skordis C.
Modified Gravity and Cosmology.
{\it Phys. Rept.}, 2012,  vol. 513, pp. 1--189.

\bibitem{Abbott:2016blz}
Abbott, B.P. et al. [LIGO Scientific and Virgo collaborations].
Observation of gravitational waves from a binary black hole merger.
{\it Phys. Rev. Lett.}, 2016, vol. 116, p. 061102.

\bibitem{cfo-Baumann:2014nda}
Baumann D.  and McAllister L.
{\it Inflation and String Theory}. Cambridge, USA: Univ. Pr., 2014, 349 p.

\bibitem{cfo-Starobinsky:1980te}
Starobinsky A. A.  A new type of isotropic cosmological models without singularity.
{\it Phys. Lett. B}, 1980, vol. 91, pp. 99--102.

\bibitem{cfo-Gottlober:1989ww}
Gottlober S., Schmidt H.J. and  Starobinsky A.A.
Sixth Order gravity and conformal transformations.
{\it Class. Quant. Grav.}, 1990, vol. 7, p. 893.

\bibitem{cfo-berkin1990effects}
Berkin A.L. and Maeda K. Effects of $R^3$ and $R\Box R$ terms on $R^2$ inflation.
{\it Phys. Lett. B}, 1990, vol. 245, pp. 348--354.

\bibitem{cfo-Amendola:1993bg}
Amendola L., Mayer B.A., Capozziello S., Occhionero F., Gottlober S., Muller V. and Schmidt H.J.
Generalized sixth order gravity and inflation.
{\it Class. Quant. Grav.}, 1993, vol. 10, L43--L47.

\bibitem{cfo-chiba2005generalized}
Chiba T.
Generalized gravity and ghost.
{\it JCAP}, 2005, vol. 0503, p. 008.

\bibitem{cfo-Cuzinatto:2013pva}
Cuzinatto R.R., de Melo C.A.M., Medeiros L.G., and Pompeia P.J.
Observational constraints on a phenomenological $f(R,\partial R)$-model.
{\it Gen. Rel. Grav.}, 2015, vol. 47, p. 29.

\bibitem{cfo-Castellanos:2018dub}
Castellanos A.R., Sobreira F., Shapiro I.L. and Starobinsky A.A.
On higher derivative corrections to the $R+R^2$ inflationary model.
{\it JCAP}, 2018, vol. 1812, p. 007.

\bibitem{cfo-BOS1992}
Buchbinder I.L., Odintsov S. and Shapiro I.
{\it Effective Action in Quantum Gravity}.
Bristol, UK: IOP Publishing, 1992, 413 p.

\bibitem{cfo-Naruko:2015zze}
Naruko A., Yoshida D. and Mukohyama S.
Gravitational scalar-tensor theory.
{\it Class. Quant. Grav.}, 2016, vol. 33, 09LT01.

\bibitem{cfo-Tsoukalas2017}
Saridakis E.N. and Tsoukalas M.
Cosmology in new gravitational scalar-tensor theories.
{\it Phys. Rev. D}, 2016,  vol. 93, p. 124032.

\bibitem{cf-CHNM2017}
Chervon S.V., Nikolaev A.V. and Mayorova T.I.
On the derivation of field equation of $f(R)$ gravity with kinetic scalar curvature.
{\it Space, Time and Fundamental Interactions}, 2017,  no. 1, pp. 30--37.

\bibitem{cfo-Chervon:2018ths}
Chervon S.V., Nikolaev A.V., Mayorova T.I., Odintsov S.D. and Oikonomou V.K.
Kinetic scalar curvature extended $f(R)$ gravity.
{\it Nucl. Phys. B}, 2018, vol. 936, pp. 597--614.

\bibitem{Chervon:2019jfu}
Chervon, S. V. and Fomin, I. V. and Mayorova, T. I.,
{\it Chiral Cosmological Model of $f(R)$ Gravity with a Kinetic Curvature Scalar}, Grav. Cosmol., v.25, No. 3, pp.205-212 (2019)
%

\bibitem{cfo-chervon2013}
Chervon S.V.
Chiral cosmological models: dark sector fields description.
{\it Quantum Matter}, 2013, vol. 2, pp. 71--82.

\bibitem{Chervon95b}
Chervon S.V.
On the chiral model of cosmological inflation.
{\it Russ. Phys. J.}, 1995, vol. 38, pp. 539--543.

\bibitem{Chervon95a}
Chervon S.V.
Chiral non-linear sigma models and cosmological inflation.
{\it Grav. Cosmol.}, 1995, vol. 1, pp. 91--96.

\bibitem{ch00mg}
Chervon S.V.
Exact solutions in standard and chiral inflationary models.
{\it Proceedings of 9th Marcell Grossman Conference, Roma, World Scientific}, 2001, p.1909.

\bibitem{ch02gc}
Chervon S.V.
A global evolution of the universe filled with scalar or chiral fields.
{\it Grav. Cosmol. Suppl.}, 2002, vol. 8, p. 32.

\bibitem{cfo-CHFK2015}
Chervon S.V., Fomin I.V., and Kubasov A.S.
{\it Scalar and Chiral Cosmological Fields}. Ulyanovsk: Ulyanovsk State Pedagogical University, 2015, 215 p.

\bibitem{bronnikov08}
Bronnikov K.A. and  Rubin S.G.
{\it Black Holes, Cosmology and Extra Dimensions}. Beijing: Natl. Ctr. Space Weather, 2012, 444 p.


\end{thebibliography}
\end{document}